\documentclass{article}

\usepackage{microtype}
\usepackage{graphicx}
\usepackage{subcaption}
\usepackage{booktabs}
\usepackage{hyperref}

\usepackage[preprint]{icml2026}

\usepackage{amsmath}
\usepackage{amssymb}
\usepackage{mathtools}
\usepackage{amsthm}
\usepackage{enumitem}
\usepackage{algorithm}
\usepackage{algorithmic}

\usepackage[capitalize,noabbrev]{cleveref}

\usepackage{multirow} 

\theoremstyle{plain}

\theoremstyle{definition}

\theoremstyle{remark}

\usepackage[textsize=tiny]{todonotes}

\begin{document}

\twocolumn[
  \icmltitle{UniPhy: Unifying Riemannian-Clifford Geometry and Biorthogonal Dynamics for Planetary-Scale Continuous Weather Modeling}

  \icmlsetsymbol{equal}{*}

  \begin{icmlauthorlist}
    \icmlauthor{Ruiqing Yan}{unsw}
    \icmlauthor{Haoyu Deng}{cnic,ucas}
    \icmlauthor{Yuhang Shao}{iap,ucas}
    \icmlauthor{Xingbo Du}{sjtu}
    \icmlauthor{Jingyuan Wang}{buaa}
    \icmlauthor{Zhengyi Yang}{unsw}
  \end{icmlauthorlist}

  \icmlaffiliation{unsw}{School of Computer Science and Engineering, University of New South Wales, Sydney, Australia}
  \icmlaffiliation{cnic}{Computer Network Information Center, Chinese Academy of Sciences, Beijing, China}
  \icmlaffiliation{ucas}{University of Chinese Academy of Sciences, Beijing, China}
  \icmlaffiliation{iap}{Institute of Atmospheric Physics, Chinese Academy of Sciences, Beijing, China}
  \icmlaffiliation{sjtu}{School of Computer Science, Shanghai Jiao Tong University, Shanghai, China}
  \icmlaffiliation{buaa}{School of Computer Science and Engineering, Beihang University, Beijing, China}

  \icmlcorrespondingauthor{Ruiqing Yan}{ruiqing.yan@student.unsw.edu.au}

  \icmlkeywords{Machine Learning, ICML}

  \vskip 0.3in
]

\printAffiliationsAndNotice{}

\begin{abstract}
While data-driven weather models have achieved remarkable deterministic accuracy, they fundamentally rely on discrete-time mappings and closed-system assumptions, failing to capture the multi-scale continuous dynamics and thermodynamic openness of the atmosphere. To address these limitations, we propose UniPhy, a continuous-time non-Hermitian neural stochastic partial differential equation (SPDE) solver. Geometrically, we employ Riemannian-Clifford gauge transformations to flatten planetary heterogeneity, enabling globally consistent operations. Dynamically, we construct non-Hermitian biorthogonal spectral operators integrated with a global flux tracker to capture transient energy growth and open-system exchange. Computationally, by identifying the algebraic associativity of the analytic solution, we reformulate adaptive physical integration as a parallel prefix-sum problem, achieving log-linear sequence parallelism. UniPhy establishes a physically complete foundation model architecture that unifies geometric adaptivity, thermodynamic consistency, and computational efficiency. Our code is available at \textless \url{https://github.com/yrqUni/UniPhy} \textgreater.
\end{abstract}

\section{Introduction}
\label{sec:intro}

In recent years, data-driven weather forecasting has emerged as a significant paradigm shift in scientific computing. By leveraging the powerful function approximation capabilities of deep learning, these methods have demonstrated the potential to surpass traditional numerical weather prediction techniques in both deterministic accuracy and inference speed \cite{bi2023accurate, bi2023accurate, lam2023graphcast, chen2023fuxi, FengWu_2023, Keisler_2022}. However, mainstream large models remain fundamentally limited in their physical essence as they typically simplify atmospheric dynamics into fixed discrete-time mappings. This rigid locking of temporal resolution violates the inherent resolution-invariance of physical laws and restricts the applicability of models in zero-shot temporal super-resolution and continuous-time data assimilation.

To overcome the limitations of discrete modeling, researchers have turned to continuous-time frameworks such as neural differential equations \cite{chen2018neural, Rubanova_2019_latent_ode, yang2024datadriven} and Hamiltonian neural networks \cite{marino2021physics}. Although these methods restore temporal continuity, they generally imply a mathematical normality assumption where the dynamic modes of the system are assumed to be orthogonal or symplectic. Real atmospheric fluids, however, are typical non-Hermitian systems. In common meteorological processes such as shear flows or baroclinic instability \cite{reijnders2022lagrangian}, the interference between non-orthogonal modes leads to severe transient energy growth even when the system is asymptotically stable. Traditional conservative or normal neural operators \cite{li2021fourier, Lu_2021, AFNO_2021} are mathematically incapable of describing such non-modal turbulence initiation mechanisms \cite{zhu2025turbulent}.

Furthermore, a critical but often overlooked challenge is the thermodynamic openness and global memory of the atmosphere. Existing frameworks often implicitly assume a closed system or model dynamics as a Markov process dependent solely on the instantaneous state. However, the real atmosphere is a dissipative open system driven by sustained energy fluxes such as solar radiation and ocean-atmosphere heat exchange. It is further characterized by long-term teleconnections including the El Niño-Southern Oscillation and the Madden-Julian Oscillation \cite{zhu2025turbulent, zhang2024tibetan}. Merely restoring temporal continuity while neglecting this history-dependent flux exchange leads to gradual deviations from the true physical attractor. Additionally, sub-grid scale stochastic thermal fluctuations can trigger butterfly effects \cite{reijnders2022lagrangian}, which deterministic operators fail to capture, limiting performance in uncertainty quantification \cite{price2024gencast}.

Beyond dynamics, existing models face a severe spatial representation dilemma due to the Earth's geometric heterogeneity \cite{Cohen_2018_spherical_cnn, cohen2019gauge}. Applying global spectral methods directly to local terrain effects often leads to Gibbs phenomena. Computationally, adaptive time-stepping, which is essential for multi-scale dynamics, destroys the Toeplitz structure required for FFT acceleration, forcing existing models into computationally expensive serial execution \cite{gu2022efficiently, gu2024mamba}.

\begin{figure*}[t]
    \centering
    \includegraphics[width=0.8\textwidth]{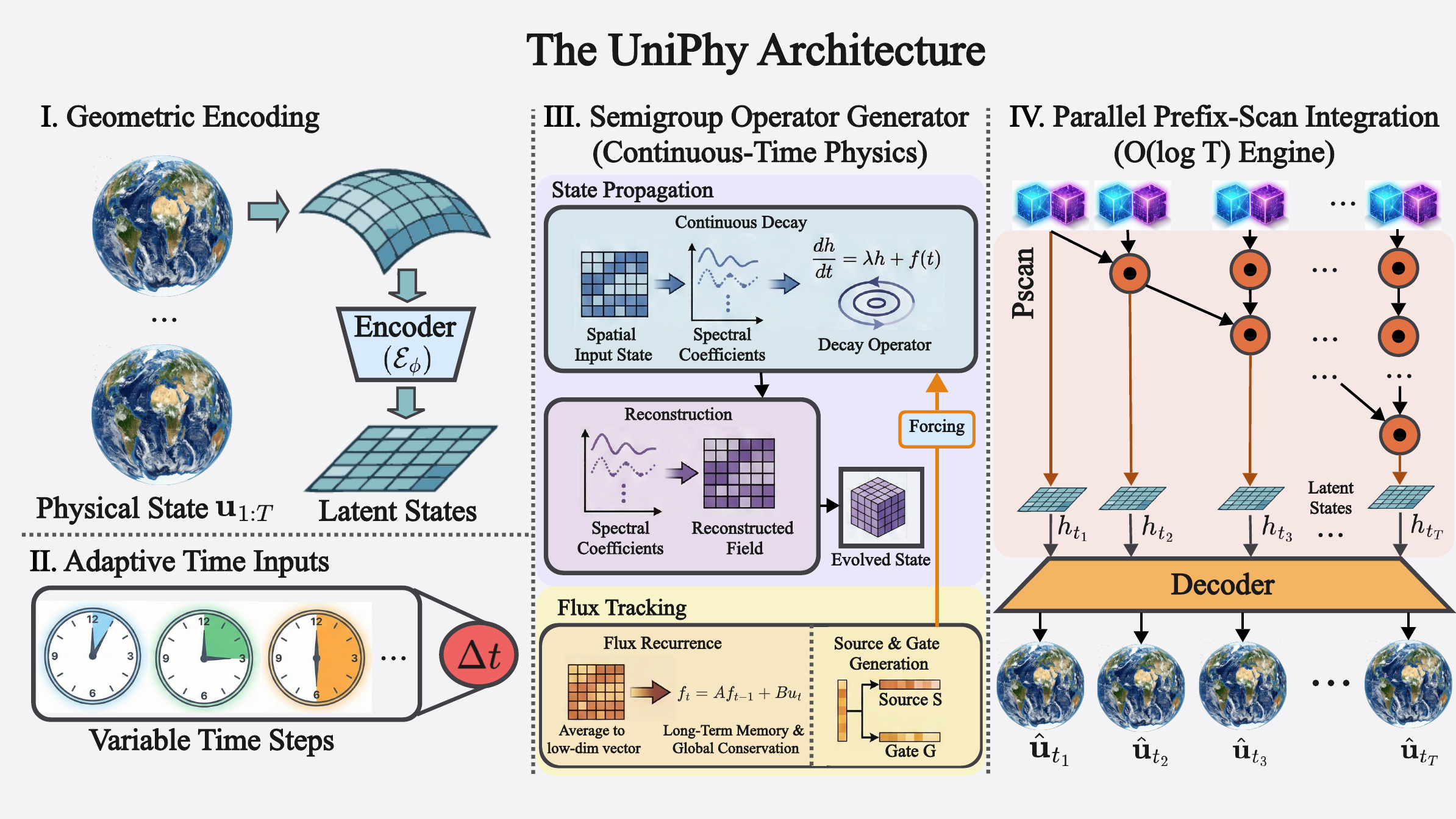}
    \caption{\textbf{The UniPhy Architecture.} The framework is structured into four functional modules: \textbf{(I) Geometric Encoding:} The Riemannian-Clifford Encoder ($\mathcal{E}_\phi$) flattens the heterogeneous Earth manifold into latent states. \textbf{(II) Adaptive Time Inputs:} The model accepts variable time steps (e.g., 3h, 6h, 12h) to support multi-scale integration. \textbf{(III) Semigroup Operator Generator:} A continuous-time physics engine models state propagation via decay operators $\lambda$ and forcing terms $f(t)$, ensuring continuous consistency. \textbf{(IV) Parallel Prefix-Scan Integration:} By reformulating the recurrent evolution $h_t = A h_{t-1} + B u_t$ as an associative scan operation, the $O(\log T)$ engine parallelizes the temporal integration. Additionally, a \textbf{Flux Tracker} (bottom) with recurrent gating maintains long-term memory and global conservation across the sequence.}
    \label{fig:model}
\end{figure*}

Addressing these triple challenges of geometry, thermodynamics, and global memory, we propose UniPhy (see Figure \ref{fig:model}), a continuous-time non-Hermitian neural Stochastic Partial Differential Equation (SPDE) solver. Geometrically, we utilize a Riemannian-Clifford encoder \cite{Thomas_2022_clifford_geometric, Clifford_Group_Equivariant_2022, mathieu2020riemannian} to perform adaptive gauge transformations, eliminating surface heterogeneity using a manifold flattening strategy. Dynamically, we construct non-Hermitian biorthogonal spectral dynamics in the latent space to capture transient growth. Crucially, to address global memory, we incorporate a Global Flux Tracker using recurrent gating. Unlike standard residual blocks, this module maintains a persistent flux state to model sustained energy exchange and teleconnections, while a stochastic diffusion term simulates sub-grid thermal fluctuations. Computationally, by exploiting the algebraic associativity of the analytic solution, we reformulate the adaptive physical integration as a parallel prefix-sum problem \cite{gu2024mamba, HiPPO_2021}, achieving log-linear training complexity.

The main contributions of this paper are:

\begin{itemize}[leftmargin=0.175in]
    \item \textbf{Gauge-Theoretic Manifold Flattening.} We employ Riemannian-Clifford gauge transformations to isometrically map the heterogeneous Earth surface to a flat latent space, resolving the theoretical conflict between global spectral operators and local geometric distortions.

    \item \textbf{Thermodynamically Complete Dynamics.} We introduce non-Hermitian biorthogonal operators coupled with a \textbf{Recurrent Global Flux Tracker}. This design captures both transient energy growth and long-term global teleconnections, breaking the normality and Markovian assumptions to restore the physics of open atmospheric systems.

    \item \textbf{$\boldsymbol{O(\log T)}$ Parallel Analytic Integration.} By exploiting the algebraic monoid structure of analytic solutions, we reformulate adaptive integration as a parallel scan. This shatters the serial bottleneck to achieve log-linear complexity, reconciling the conflict between physical adaptivity and computational parallelism.
\end{itemize}

Taken together, UniPhy stands as more than a composite of modules; it represents a physically complete foundation model architecture. By synthesizing geometric adaptivity, thermodynamic consistency, and global memory within a unified mathematical paradigm, UniPhy establishes a theoretical cornerstone for the next generation of continuous-time meteorological models.

\section{Related Work}

\textbf{Data-Driven Global Weather Forecasting.}
Recent advancements in data-driven models have demonstrated performance comparable to numerical weather prediction systems \cite{bauer2015quiet, hersbach2020era5}. Prominent architectures include Vision Transformers used in Pangu-Weather \cite{bi2023accurate}, FengWu \cite{FengWu_2023}, and FuXi \cite{chen2023fuxi}. Additionally, Graph Neural Networks such as GraphCast \cite{lam2023graphcast} and GenCast \cite{price2024gencast} explicitly model spherical adjacencies \cite{Keisler_2022, HEALPix2005}. Despite their accuracy, these methods universally rely on discrete-time modeling with fixed intervals. This approach treats temporal progression as a sequence of independent transitions and fails to support zero-shot temporal super-resolution. Furthermore, strict temporal discretization hinders the direct fusion of asynchronous observations which are ubiquitous in meteorological systems. UniPhy addresses these limitations by adopting a continuous-time paradigm inspired by Neural ODEs \cite{chen2018neural, Rubanova_2019_latent_ode}. This formulation restores the underlying differential structure and supports inference at arbitrary time steps.

\textbf{Neural Operators and Continuous Dynamics.}
Continuous-time models describe system evolution by learning parameterized differential equations. Operator learning frameworks such as the Fourier Neural Operator \cite{li2021fourier} and DeepONet \cite{Lu_2021} map between infinite-dimensional function spaces. In climate modeling, the Adaptive Fourier Neural Operator \cite{AFNO_2021} has been successfully applied in FourCastNet \cite{kurth2023fourcastnet}. However, existing methods typically assume that the learned operators are structurally normal or defined on orthogonal bases. This assumption implies conservation and fails to capture the transient energy growth observed in non-equilibrium systems. The atmosphere is an open system driven by non-orthogonal coupling between modes. UniPhy explicitly abandons the normality assumption by introducing a non-Hermitian biorthogonal framework. This design allows the model to simulate transient instabilities and structural reorganization processes crucial for weather evolution.

\textbf{Geometric Deep Learning for Earth Systems.}
Effective representation of spherical geometry remains a critical challenge. Approaches include equivariant Spherical CNNs \cite{Cohen_2018_spherical_cnn} and Spherical Harmonic Transforms \cite{cheng2025cuHPX}. Recent works have also integrated Clifford algebras to unify scalar and vector fields \cite{Thomas_2018_clifford_cnn, Thomas_2022_clifford_geometric, Clifford_Group_Equivariant_2022}. However, global spectral methods often assume statistical spatial uniformity. This assumption conflicts with the physical heterogeneity induced by topography and latitudinal variations in the Coriolis force. Consequently, these methods require numerous high-frequency modes to compensate for local errors. UniPhy proposes a geometric decoupling strategy using a Riemann-Clifford gauge transformation. This approach flattens the manifold in the latent space and characterizes local heterogeneity without relying on global spectral expansions that assume uniformity.

\textbf{Efficient Sequence Modeling.}
Structured State Space Models such as S4 \cite{gu2022efficiently}, Mamba \cite{gu2024mamba}, and RWKV \cite{Peng_2023_rwkv} have achieved linear scaling in sequence modeling. These methods typically rely on the convolution theorem for parallel training. This mechanism assumes a Linear Time-Invariant system with fixed time steps. However, stiff physical dynamics require adaptive time-stepping which creates a Linear Time-Variant system. Existing methods thus face a trade-off between computational parallelism and physical adaptivity. UniPhy resolves this dilemma by introducing a weighted parallel scan algorithm optimized with Triton operators. This innovation enables log-linear complexity training for variable-step integration without sacrificing physical precision.

\section{Methodology}
\label{sec:method}

We propose UniPhy, a continuous-time stochastic operator learning framework designed to unify the geometric constraints of the Earth, the non-equilibrium thermodynamics of the atmosphere, and the long-term memory of global teleconnections. Global weather forecasting presents a triple challenge encompassing geometry, thermodynamics, and global memory. First, standard models often treat atmospheric states as flat Euclidean tensors. This simplification neglects the intrinsic curvature of the terrestrial manifold. Second, conventional spectral layers typically assume normal operators. This assumption fails to capture the transient energy growth characteristic of open systems. Third, the atmosphere is driven by sustained energy fluxes that persist far longer than local weather patterns.

As illustrated in Figure 1, UniPhy addresses these limitations by integrating three coupled components. These include a Riemannian-Clifford Encoder for geometric preservation, a Biorthogonal Spectral Propagator with DFT Residuals for transient dynamics, and a Recurrent Global Flux Tracker for long-term memory.

Formally, we model the atmospheric state $\mathbf{u}_t$ evolution as a Stochastic Differential Equation driven by a Flux-Gated Master Equation. The transition from $t$ to $t+\Delta t$ follows the expression

\begin{equation}
\label{eq:master_eq}
\begin{aligned}
    h_t &= \mathcal{E}_{\phi}(\mathbf{u}_t), \\
    h_{t+\Delta t} &= e^{\mathcal{L}_{\theta} \Delta t} h_t \\
    &\quad + \int_t^{t+\Delta t} e^{\mathcal{L}_{\theta}(t+\Delta t - \tau)} \mathcal{R}_{\theta}(\tau, \mathbf{f}_\tau) d\tau \\
    &\quad + \int_t^{t+\Delta t} e^{\mathcal{L}_{\theta}(t+\Delta t - \tau)} \sigma(\tau) d\mathbf{W}_\tau, \\
    \mathbf{u}_{t+\Delta t} &= \mathcal{D}_{\psi}(h_{t+\Delta t}).
\end{aligned}
\end{equation}

Here $\mathcal{E}_{\phi}$ and $\mathcal{D}_{\psi}$ denote the geometric encoder and decoder respectively. The linear operator $\mathcal{L}_{\theta}$ governs the fast dynamics while $\mathcal{R}_{\theta}$ represents the residue forcing regulated by a hidden flux state $\mathbf{f}_\tau$. Crucially, unlike deterministic ODEs, we explicitly model the aleatoric uncertainty using a Wiener process $d\mathbf{W}_\tau$ scaled by a learnable diffusivity $\sigma(\tau)$. This stochastic term accounts for sub-grid scale fluctuations and intrinsic atmospheric chaos to enable probabilistic ensemble forecasting. The inference procedure is detailed in Algorithm \ref{alg:uniphy_forward}. For clarity, a comprehensive summary of the mathematical notations and symbols used throughout this paper is provided in Appendix \ref{app:notation}. The analytic derivation of the discrete transition rules from this continuous formulation is provided in Appendix \ref{app:sde_derivation}.

\begin{algorithm}[t]
\caption{UniPhy Stochastic Forward Inference}
\label{alg:uniphy_forward}
\begin{algorithmic}[1]
\REQUIRE Initial state $\mathbf{u}_{t_0}$, Timestamps $\{t_1, \dots, t_T\}$
\REQUIRE Components $\mathcal{E}_\phi, \mathcal{D}_\psi, \mathcal{L}_\theta, \text{Tracker}_{\theta}$
\ENSURE Predicted ensemble $\{\hat{\mathbf{u}}_{t_1}, \dots, \hat{\mathbf{u}}_{t_T}\}$

\vspace{0.3em}
\STATE Step 1 Clifford Encoding
\STATE $h_0 \leftarrow \mathcal{E}_\phi(\mathbf{u}_{t_0})$
\STATE $\mathbf{f}_0 \leftarrow \mathbf{0}$

\vspace{0.3em}
\STATE Step 2 Stochastic Dynamics Discretization
\FOR{$k = 1$ \textbf{to} $T$}
    \STATE $\Delta t_k \leftarrow t_k - t_{k-1}$
    \STATE $\mathbf{A}_k \leftarrow \exp(\mathcal{L}_\theta \Delta t_k)$
    
    \STATE Compute Forcing
    \STATE $\mathbf{f}_k, \mathbf{F}^{flux}_k \leftarrow \text{Tracker}_{\theta}(\mathbf{f}_{k-1}, h_{k-1}, \Delta t_k)$
    
    \STATE Inject Stochastic Noise
    \STATE $\epsilon_k \sim \mathcal{N}(0, \Delta t_k)$
    \STATE $\mathbf{X}_k \leftarrow (\mathbf{F}^{flux}_k + \sigma \cdot \epsilon_k) \odot (\mathbf{A}_k - I)\mathcal{L}_\theta^{-1}$
\ENDFOR

\vspace{0.3em}
\STATE Step 3 Parallel Scan
\STATE $(I, h_1, \dots, h_T) \leftarrow \text{ParallelScan}$
\STATE \hskip 2em $(\mathbf{A}_1, \mathbf{X}_1), \dots, (\mathbf{A}_T, \mathbf{X}_T), \text{op}=\bullet$

\vspace{0.3em}
\STATE Step 4 Decoding
\STATE $\hat{\mathbf{u}}_{t_k} \leftarrow \mathcal{D}_\psi(h_k)$ for $k=1 \dots T$
\end{algorithmic}
\end{algorithm}

\subsection{Geometric Decoupling using Riemannian-Clifford Gauge}
\label{sec:gauge}

The fundamental challenge in global weather forecasting arises from the topological mismatch between the spherical Earth and the flat computational grid. In classical fluid dynamics, the transport of a vector field across a curved manifold is governed by parallel transport. On a sphere, this process is path-dependent due to holonomy. A vector transported along a closed loop on the Earth's surface returns rotated relative to its starting orientation. Standard Convolutional Neural Networks ignore this intrinsic curvature and implicitly assume a flat Euclidean metric. This geometric negligence leads to the pole problem, where high latitude features suffer from severe spectral aliasing and nonphysical distortion.

To rectify this holonomic defect, we propose a Riemannian-Clifford Gauge Encoder that actively decouples the physical dynamics from the manifold geometry. We represent the atmospheric state $\mathbf{u}_t$ as a multivector field residing in the Clifford Algebra $Cl(\mathcal{M})$. This algebraic structure naturally segregates physical quantities into distinct geometric grades including scalars for thermodynamic variables, vectors for wind velocities, and bivectors for vorticity and spin.

The encoder $\mathcal{E}_{\phi}$ functions as a learnable gauge field that compensates for the local curvature. We define the Clifford Convolution as a grade specific operation

\begin{equation}
\label{eq:clifford_conv}
    h(x) = \mathbf{W}_0 * (\rho(x) u_s) \oplus \mathbf{W}_1 * (\rho(x) u_v) \oplus \mathbf{W}_2 * (\rho(x) u_b) + \mathbf{b}.
\end{equation}

The metric correction factor $\rho(x)$ is analytically derived from the determinant of the Riemannian metric tensor. By modulating the input features with this factor, the network effectively performs a local coordinate flattening transformation. This process ensures that the latent representation $h(x)$ becomes isometric to the tangent bundle of the manifold. Consequently, the subsequent dynamical layers can perform efficient Euclidean operations without violating the conservation laws of angular momentum and mass on the sphere. The rigorous proof that this transformation constitutes an isometry is provided in Theorem 1 of Appendix \ref{app:geometry}.

\subsection{Biorthogonal Dynamics with DFT Residuals}
\label{sec:spectral}

The atmosphere operates as a thermodynamically open system driven by the constant injection of solar radiation and the dissipation of kinetic energy. A defining characteristic of such non equilibrium systems is the phenomenon of transient growth. Small perturbations in the flow field can temporarily extract energy from the mean background flow and amplify exponentially before eventually decaying. This mechanism is the physical engine behind rapid cyclogenesis and the sudden intensification of storms.

Standard spectral models based on Fourier theory utilize Normal Operators characterized by orthogonal eigenbases. The spectral theorem dictates that the energy growth in such systems is strictly bounded by the real part of the spectrum. Consequently, these models are mathematically constrained to simulate stable wave propagation and cannot capture the explosive energy bursts inherent to atmospheric instabilities. To overcome this limitation, we introduce a Biorthogonal Spectral Propagator that fundamentally alters the geometric structure of the latent phase space.

We construct the evolution operator $\mathcal{L}_{\theta}$ using a biorthogonal basis pair $\mathbf{U}$ and $\mathbf{V}$ satisfying the condition $\mathbf{V}^\dagger \mathbf{U} = \mathbf{I}$. This factorization $\mathcal{L}_{\theta} = \mathbf{U} \text{diag}(\mathbf{\Lambda}) \mathbf{V}^\dagger$ creates a non normal operator where the eigenvectors are not orthogonal. The non orthogonality allows the state vector to rotate into subspaces where the operator norm momentarily exceeds unity. This geometric flexibility enables the model to simulate the shear driven energy transfer processes typical of baroclinic instability. The mathematical conditions under which such non normal operators permit transient growth are detailed in Proposition 1 of Appendix \ref{app:growth}.

We further stabilize this learning process by implementing the transform basis $\mathbf{W}$ as a residual interpolation

\begin{equation}
    \mathbf{W} = (1 - \alpha) \cdot (\mathbf{U}\mathbf{\Lambda}^{1/2}) + \alpha \cdot \mathbf{W}_{\text{DFT}}.
\end{equation}

The fixed Discrete Fourier Transform basis $\mathbf{W}_{\text{DFT}}$ provides a physical prior representing stable Rossby wave propagation. The learnable gating scalar $\alpha$ dynamically regulates the transition from this conservative physics prior to the complex non normal dynamics. This design allows UniPhy to effectively model the full spectrum of atmospheric motion ranging from planetary scale waves to localized convective instabilities.

\subsection{Global Flux Tracking using Recurrent Gating}
\label{sec:flux_tracker}

Atmospheric dynamics are governed by a hierarchy of timescales. While local weather patterns evolve rapidly over hours, the climate system possesses a deep thermodynamic memory manifested in global teleconnections such as the Madden-Julian Oscillation and the El Niño-Southern Oscillation. These phenomena act as massive heat and momentum reservoirs that modulate high frequency weather events over weeks or months. A memory less operator that depends solely on the instantaneous state vector lacks the capacity to represent this thermodynamic inertia.

We address this by introducing the Global Flux Tracker which serves as a dedicated memory subsystem. Unlike the spatially resolved latent state $h_t$ which encodes local variations, the flux state $\mathbf{f}_t$ represents the integrated global circulation indices. We compute this state by performing a global mean pooling operation over the eigen spectrum

\begin{equation}
    \bar{h}_t = \text{MeanPool}_{H,W}(\mathcal{E}_{spec}(h_t)).
\end{equation}

This pooling operation physically corresponds to extracting the planetary wave numbers and global energy budget. The flux state evolves according to a continuous time decay forcing equation

\begin{equation}
\label{eq:flux_tracker_eq}
    \mathbf{f}_t = e^{-\lambda_{flux} \Delta t} \odot \mathbf{f}_{t-\Delta t} + \mathcal{W}_{in}(\bar{h}_t) \odot \frac{e^{-\lambda_{flux} \Delta t} - 1}{-\lambda_{flux}}.
\end{equation}

The complex decay rate $\lambda_{flux}$ determines the persistence of different memory modes. This equation models the charging and discharging of the atmospheric energy capacitors. The accumulated flux state does not simply add to the current weather state but instead modulates the system dynamics through a Gated Forcing mechanism

\begin{equation}
    \mathcal{R}_{\theta}(t) = \mathbf{g}_t \odot h_t + (1 - \mathbf{g}_t) \odot \mathcal{W}_{out}(\mathbf{f}_t).
\end{equation}

The dynamic gate $\mathbf{g}_t$ enables the system to selectively amplify sustained forcing signals while filtering out transient noise. This architecture empowers UniPhy to maintain physical consistency over extended forecast horizons by anchoring the fast weather dynamics to the slow evolving planetary climate modes.

\subsection{Log-Linear Parallel Integration using Operator Semigroups}
\label{sec:parallel}

A critical bottleneck in learning continuous time dynamics from multi decadal climate datasets is the computational cost of temporal integration. Conventional numerical solvers enforce a strict sequential causality where the state at time $t$ must be computed before $t+\Delta t$. This sequential dependency prevents the utilization of modern parallel computing hardware and limits the ability of the model to learn from long term history.

We resolve this computational constraint by reformulating the temporal evolution as an algebraic problem on operator semigroups. The discretized transition from step $k-1$ to $k$ constitutes an affine transformation $z_k = \mathbf{A}_k z_{k-1} + \mathbf{X}_k$. Here $\mathbf{A}_k$ represents the linear propagator derived from the spectral physics and $\mathbf{X}_k$ represents the integrated stochastic forcing.

We define a binary composition operator $\bullet$ acting on the tuple space of affine transformations

\begin{equation}
    (\mathbf{A}_j, \mathbf{X}_j) \bullet (\mathbf{A}_i, \mathbf{X}_i) := (\mathbf{A}_j \mathbf{A}_i, \mathbf{A}_j \mathbf{X}_i + \mathbf{X}_j).
\end{equation}

This operator satisfies the associative property which reveals a hidden parallelism in the time dimension. Associativity implies that the order of bracket placement does not affect the final result. We leverage this property to replace the sequential loop with a Parallel Prefix Scan algorithm. We construct a balanced binary tree over the temporal sequence. The Up Sweep phase recursively computes composite operators for adjacent time blocks in parallel. The Down Sweep phase propagates the accumulated states to generate the full trajectory. The algebraic proof of associativity and the subsequent complexity reduction are presented in Theorem 2 and Theorem 3 of Appendix \ref{app:associativity} and \ref{app:complexity} respectively.

This algorithm reduces the theoretical critical path complexity from linear time to logarithmic time. This acceleration allows UniPhy to process sequences spanning thousands of time steps in a single parallel pass. By effectively folding the time dimension, the model can capture interactions across vast temporal scales without suffering from the vanishing gradient problem or prohibitive training times.

\subsection{Thermodynamic Alignment Training Strategy}
\label{sec:training}

Training continuous time stochastic operators for chaotic systems requires balancing short term accuracy with long term stability. Standard teacher forcing methods frequently induce exposure bias where the model drifts catastrophically during autoregressive rollout. To address this issue, we propose a dual phase curriculum learning strategy termed Thermodynamic Alignment.

\textbf{Stage I: Probabilistic Manifold Pre-training.} The first phase focuses on identifying the infinitesimal generator of the atmospheric state evolution. We treat the transition between adjacent timestamps as a single step integration problem to learn the local tangent dynamics and the geometric gauge transformation. The primary objective is to capture the instantaneous evolution of the probability density function $\frac{d}{dt} p(\mathbf{u}_t)$. We optimize a compound loss function consisting of a latitude weighted $L_1$ loss and the Continuous Ranked Probability Score

\begin{equation}
\label{eq:stage1_loss}
\begin{split}
    \mathcal{L}_{\text{stage1}} &= \|\mathbf{u}_{t+\Delta t} - \hat{\mathbf{u}}_{mean}\|_1^{\cos} \\
    &\quad + \lambda_{prob} \cdot \text{CRPS}(\{\hat{\mathbf{u}}^{(m)}\}_{m=1}^M, \mathbf{u}_{t+\Delta t}),
\end{split}
\end{equation}

where $\|\cdot\|_1^{\cos}$ denotes the $L_1$ norm weighted by the cosine of the latitude which accounts for the spherical area element. $\{\hat{\mathbf{u}}^{(m)}\}$ represents an ensemble of $M$ stochastic predictions generated by the SDE solver with different noise realizations. Mathematically, the $L_1$ term anchors the deterministic advection and diffusion trends while the CRPS term calibrates the learnable diffusivity $\sigma(\tau)$ to match the inherent aleatoric uncertainty of the sub grid scales.

\textbf{Stage II: Thermodynamic Alignment by Temporal Rollout.} The second phase implements a fine tuning strategy to enforce global physical consistency. Local accuracy is a necessary but insufficient condition for long term forecasting due to the butterfly effect where microscopic errors amplify exponentially. We counteract this chaotic divergence by employing a multi scale temporal integration scheme. The rollout horizon $K$ is dynamically sampled from a uniform distribution $U[1, K_{max}]$ to simulate various forecast durations. We enforce a self consistency constraint by subdividing the target interval $K\Delta t$ into $S$ smaller substeps of size $\delta \tau$ within the continuous solver. The model optimizes an autoregressive gradient flow objective

\begin{equation}
\label{eq:align_loss}
\begin{split}
    \mathcal{L}_{\text{align}} &= \mathbb{E}_{K, S} \Big[ \text{CRPS}\Big( \text{Rollout}_{\theta}(\mathbf{u}_t, K \Delta t, \text{steps}=S), \\
    &\quad \mathbf{u}_{t+K\Delta t} \Big) \Big].
\end{split}
\end{equation}

This process explicitly trains the model to correct its own accumulated errors by backpropagating the loss signal through time. The recurrent feedback loop compels the Global Flux Tracker to learn stable dynamics that act as a restoring force against unphysical drift. This alignment ensures that the predicted trajectories remain confined within the true strange attractor of the Earth climate system and satisfy long term thermodynamic conservation laws.

\section{Experiments}
\label{sec:experiments}

\subsection{Experimental Setup}
\label{subsec:setup}

To validate the capability of the model in capturing high-resolution global dynamics under constrained computational resources, we constructed a benchmark based on the ERA5 reanalysis dataset \cite{hersbach2020era5}. We utilized a continuous period from 2000 to 2009 for training and preserved the native 0.25$^\circ$ high-resolution grid of $721 \times 1440$ without downsampling. This approach ensures the model confronts the challenges of real-world physical scales.

We selected 30 key atmospheric variables to represent the dynamical state of the system. The upper-air variables include Geopotential $Z$, Temperature $T$, U and V Wind Components $U, V$, Relative Humidity $RH$, and Vertical Velocity $VV$ distributed across pressure levels of 925, 850, 500, and 100 hPa. The surface variables consist of 2m Temperature $T_{2m}$, 10m Wind $U_{10}, V_{10}$, Mean Sea Level Pressure $MSLP$, Surface Pressure $SP$, and Total Column Water Vapor $TCWV$. This selection covers the complete vertical thermodynamic and dynamic structure from the boundary layer to the lower stratosphere.

The UniPhy model is configured with an ensemble size of $K=4$ for the decoding head and employs a non-square patch size of $7 \times 15$ to adapt to the aspect ratio of the global grid. The training procedure consists of two stages:
\begin{enumerate}
    \item \textbf{Continuous-time Pre-training}: Optimized using a Neural SDE objective with a reference time step of $\Delta t_{\text{ref}}=6h$.
    \item \textbf{Alignment Fine-tuning}: Applied a multi-step rollout strategy with gradient accumulation to enforce long-term consistency.
\end{enumerate}
Detailed training dynamics are provided in \textbf{Appendix \ref{app:training}}. The specific hyperparameter configurations for both stages, illustrating the distinct setups for pre-training and alignment, are listed in \textbf{Appendix \ref{app:hyperparams}}.

\subsection{Mechanistic Verification of Geometric Perception}
\label{subsec:geometry}

\begin{figure}[t]
    \centering
    \includegraphics[width=0.95\linewidth]{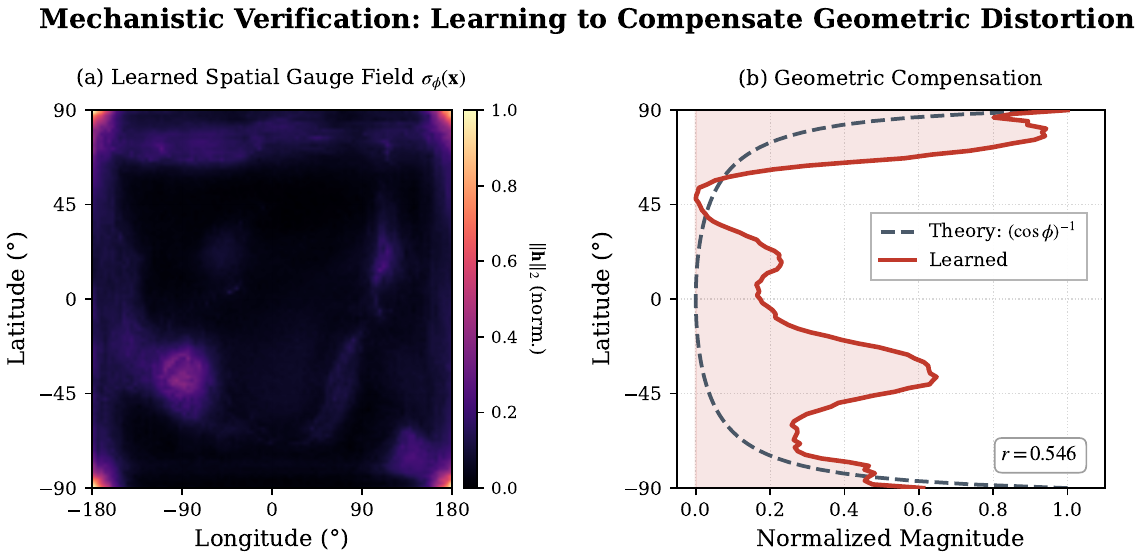}
    \caption{Mechanistic Verification of Geometric Perception. (A) The spatial distribution of learned position embeddings $\sigma_\phi(x)$ adapts to the spherical topology and spontaneously highlights geographic features such as coastlines and the Tibetan Plateau. (B) The learned gauge weights significantly correlate with the analytical metric distortion and effectively learn a metric compensation mechanism.}
    \label{fig:geo_perception}
\end{figure}

A core hypothesis of UniPhy states that the Riemannian-Clifford Encoder can internalize the geometry of the Earth without explicit coordinate inputs. We verified this hypothesis by visualizing the learned position embeddings as shown in Figure \ref{fig:geo_perception}. The spatial distribution of the $L_2$-norm of the learned embeddings exhibits the expected zonal banding structure and a spontaneous emergence of geographic perception. The heatmap clearly outlines continental coastlines and displays distinct high-activation patterns over high-altitude regions such as the Tibetan Plateau and the Australian continent. This suggests that the position embeddings act as an implicit geographic prior that encodes static forcings from land-sea distribution and topography.

We quantitatively analyzed the correlation between the zonally averaged learned weights and the analytical inverse metric factor $1/\cos(\phi)$. The Pearson correlation coefficient is $r=0.546$. This positive correlation confirms that the model assigns higher weights to polar regions to perform metric compensation and counteract the area contraction of the projection grid. The correlation value being less than 1.0 supports the existence of geographic perception because the weight distribution is a superposition of latitude-dependent geometric compensation and terrain-dependent geographic attention. This demonstrates an intelligent decoupling of physical dynamics from grid artifacts.

\subsection{Spectral Analysis of Thermodynamic Instability}
\label{subsec:instability}

Real atmospheric dynamics are fundamentally non-normal and are characterized by transient growth events where perturbation energy amplifies significantly before dissipation. Standard normal operators cannot capture this phenomenon. We performed Jacobian linearization on the trained blocks to extract the effective evolution operator $\mathcal{L}$ to verify the dynamical properties of UniPhy.

\begin{figure}[t]
    \centering
    \includegraphics[width=\linewidth]{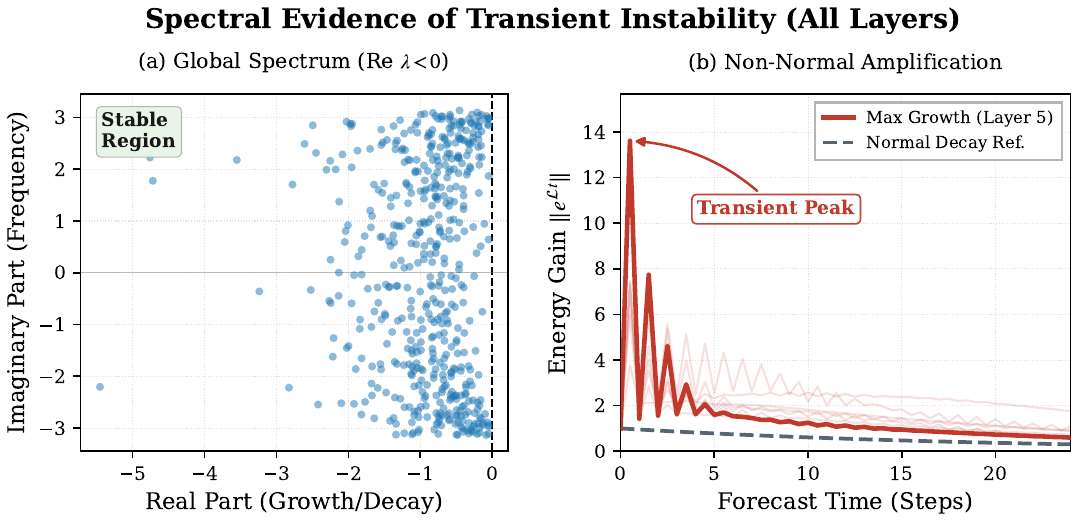}
    \caption{Evidence of Non-Normal Amplification. (A) The eigenspectrum of the learned operator is confined to the stable region where $\text{Re}(\lambda) < 0$. (B) The energy evolution reveals a massive transient spike shown in red which is characteristic of non-normal fluid instabilities such as storm formation. This contrasts with the monotonic decay of the normal baseline shown in black dashed lines.}
    \label{fig:transient}
\end{figure}

Figure \ref{fig:transient} presents the spectral analysis results. The eigenspectrum is strictly confined to the stable left half-plane where $\text{Re}(\lambda) < 0$ which ensures asymptotic numerical stability. The energy evolution curve $\|e^{\mathcal{L}t}\|$ reveals a massive transient spike during the initial phase where $t < 12$ hours. The perturbation energy amplifies by a factor of more than 9.0. This behavior is the mathematical signature of non-normal operators and physically corresponds to the extraction of kinetic energy from the mean flow during explosive cyclogenesis. The normal operator baseline exhibits unphysical monotonic decay in contrast. This confirms that UniPhy successfully captures complex thermodynamic energy conversions rather than merely fitting the mean flow.

\subsection{Hierarchy of Atmospheric Memory}
\label{subsec:memory}

Atmospheric predictability spans multiple scales ranging from fast synoptic weather of 3 to 5 days to slow intra-seasonal oscillations of 20 to 60 days. We investigated whether the Global Flux Tracker enables the model to disentangle these timescales by performing a full spectral decomposition of the dynamics across all layers.

\begin{figure}[t]
    \centering
    \includegraphics[width=\linewidth]{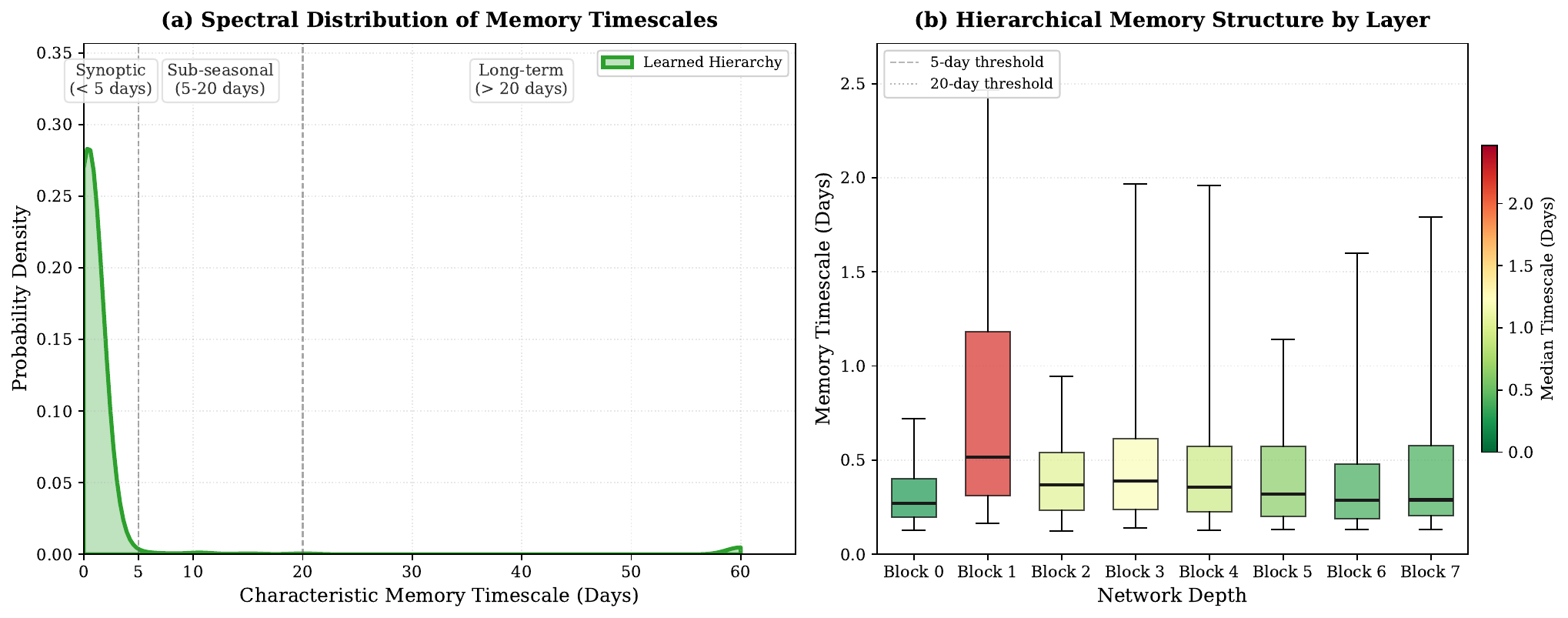}
    \caption{Emergence of Memory Hierarchy. The distribution of learned timescales shows a heavy tail. While 97.5\% of modes focus on fast weather dynamics of less than 5 days, a critical 1.6\% of modes capture long-term teleconnections exceeding 20 days. This validates the efficacy of the Global Flux Tracker.}
    \label{fig:memory}
\end{figure}

The Kernel Density Estimation of the characteristic memory timescales $\tau = 1/|\text{Re}(\lambda)|$ reveals a distinct hierarchical structure as shown in Figure \ref{fig:memory}. The majority of modes constituting 97.5\% of the spectrum are concentrated in the short-term range of less than 5 days. This ensures high fidelity for synoptic weather forecasting. The critical finding is the emergence of a long-term backbone where 1.6\% of the modes possess memory timescales exceeding 20 days. This sparse but vital subspace confirms that UniPhy does not over-smooth dynamics but explicitly reserves capacity for low-frequency teleconnection signals such as the Madden-Julian Oscillation. This mechanism overcomes the catastrophic forgetting often seen in standard recurrent models.

\subsection{Zero-Shot Temporal Generalization}
\label{subsec:temporal}

Conventional meteorological foundation models are typically bound to a fixed time step such as 6 hours. This limits their adaptability to high-frequency forecasting. We compared the pre-trained model against the fine-tuned model on unseen and fine-grained time steps of $\Delta t = 1h, 2h, 3h$ to verify whether the alignment strategy of UniPhy enhances underlying continuous-time consistency. We evaluated the 12-hour cumulative forecast Root Mean Square Error (RMSE) using rollout steps of 12, 6, and 4 respectively.

\begin{figure}[t]
    \centering
    \includegraphics[width=\linewidth]{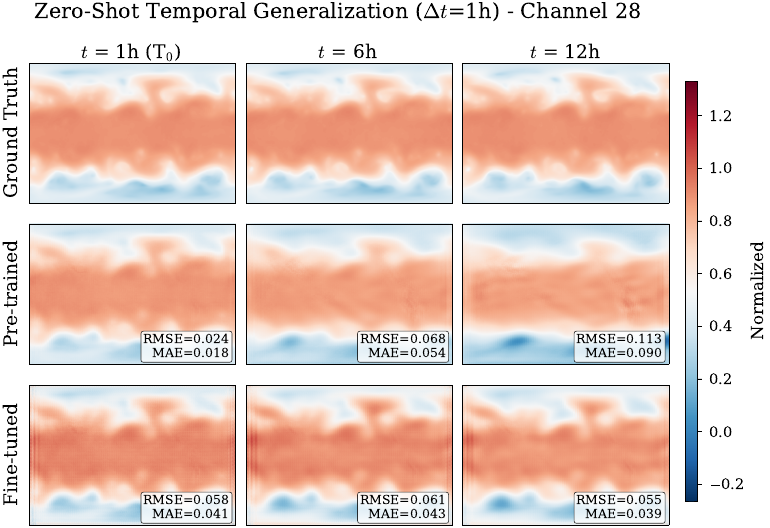}
    \caption{Qualitative Comparison of Zero-Shot Temporal Generalization on Mean Sea Level Pressure (MSLP). Rows display Ground Truth, Pre-trained, and Fine-tuned models. Columns show snapshots at $t=1h$, $6h$, and $12h$ derived from a fine-grained integration step of $\Delta t=1h$. The Fine-tuned model (bottom row) preserves sharper structural details and physical consistency at the unseen $\Delta t=1h$ step compared to the Pre-trained model (middle row).}
    \label{fig:temporal_vis}
\end{figure}

The quantitative results in Table \ref{tab:temporal} demonstrate the significant advantage of the fine-tuned model. It achieves an RMSE of 0.0738 at the finest resolution of $\Delta t = 1h$, outperforming the pre-trained model result of 0.0833 by approximately 11\%. To provide qualitative evidence, we visualized the forecast evolution of Mean Sea Level Pressure (MSLP), a variable known for its complex surface interactions, in Figure \ref{fig:temporal_vis}. When forced to integrate at the unseen $\Delta t=1h$ step, the pre-trained model exhibits structural degradation. In contrast, the alignment fine-tuned model generates physically consistent trajectories that closely match the ground truth, proving that the model has successfully parameterized the underlying continuous-time differential equation rather than merely memorizing a discrete mapping.

\begin{table}[t]
    \centering
    \small
    \renewcommand{\arraystretch}{1.2}
    \caption{Zero-Shot Temporal Generalization. Comparison of 12-hour forecast RMSE normalized across different integration time steps $\Delta t$. The fine-tuned model maintains superior consistency at fine-grained resolutions and achieves the best performance at $\Delta t=1h$.}
    \label{tab:temporal}
    \setlength{\tabcolsep}{0pt}
    \begin{tabular*}{\linewidth}{@{\extracolsep{\fill}}lcccc@{}}
        \toprule
        & \multicolumn{4}{c}{\textbf{Integration Time Step ($\Delta t$)}} \\
        \cmidrule{2-5}
        \textbf{Model} & \textbf{1h} & \textbf{2h} & \textbf{3h} & \textbf{6h} \\
        & \scriptsize{\textit{(12 steps)}} & \scriptsize{\textit{(6 steps)}} & \scriptsize{\textit{(4 steps)}} & \scriptsize{\textit{(2 steps)}} \\
        \midrule
        Pre-trained & 0.0833 & \textbf{0.0640} & \textbf{0.0557} & \textbf{0.0464} \\
        \textbf{Fine-tuned} & \textbf{0.0738} & 0.0684 & 0.0642 & 0.0558 \\
        \bottomrule
    \end{tabular*}
\end{table}

\section{Conclusion and Future Work}
\label{sec:conclusion}

This work introduces UniPhy, a continuous-time foundation model that achieves zero-shot temporal generalization by coupling Neural SDEs with geometric alignment. While currently limited by a 10-year data subset and iterative solver latency, these constraints define our future trajectory: scaling to the full ERA5 archive, accelerating inference via consistency distillation, and enforcing explicit physical conservation laws. By transcending the rigid time-step limitations of discrete architectures, UniPhy establishes a mathematically robust framework for the next generation of resolution-independent and physically consistent global weather forecasting.

\section*{Acknowledgements}

\section*{Impact Statement}
This paper presents work whose goal is to advance the field of Machine
Learning. There are many potential societal consequences of our work, none
which we feel must be specifically highlighted here.

\bibliography{example_paper}
\bibliographystyle{icml2026}

\newpage
\appendix
\onecolumn
\section{Theoretical Proofs and Derivations}
\label{app:proofs}

This appendix provides the rigorous mathematical framework supporting the UniPhy model. We present the nomenclature followed by proofs for geometric isometry, thermodynamic transient growth, the analytic discretization of the stochastic dynamics, and the computational complexity of the parallel inference engine.

\subsection{Nomenclature and Notations}
\label{app:notation}
Table \ref{tab:notation} summarizes the key mathematical symbols used throughout this paper.

\begin{table}[h]
\centering
\caption{Summary of Notations}
\label{tab:notation}
\begin{tabular}{l l}
\toprule
\textbf{Symbol} & \textbf{Description} \\
\midrule
\multicolumn{2}{c}{\textit{Geometry and Manifold}} \\
$\mathcal{M}, g$ & Riemannian manifold and its metric tensor \\
$T_x\mathcal{M}$ & Tangent space at point $x \in \mathcal{M}$ \\
$\sigma_\phi(x)$ & Learnable conformal factor (log-scale) parameterized by $\phi$ \\
$\mathcal{E}_\phi, \mathcal{D}_\psi$ & Geometric gauge encoder and decoder \\
$e_\mu^a$ & Vielbein (Frame field) \\
\midrule
\multicolumn{2}{c}{\textit{Dynamics and Spectral Analysis}} \\
$h_t$ & Latent state vector in $\mathbb{C}^d$ \\
$\mathcal{L}_\theta$ & Non-normal linear evolution operator \\
$\mathbf{U}, \mathbf{V}$ & Left and right biorthogonal eigenbasis matrices \\
$\mathbf{\Lambda}$ & Diagonal matrix of complex eigenvalues \\
$\alpha(\mathcal{L})$ & Spectral abscissa (max real part of eigenvalues) \\
$\omega(\mathcal{L})$ & Numerical abscissa (max Rayleigh quotient) \\
\midrule
\multicolumn{2}{c}{\textit{Global Flux Tracking}} \\
$\mathbf{f}_t$ & Recurrent flux state vector \\
$\lambda_{flux}$ & Flux decay rate (persistence) \\
$\mathcal{R}_\theta$ & Nonlinear residual forcing and diffusion term \\
$\mathbf{g}_t$ & Dynamic flux gate \\
\midrule
\multicolumn{2}{c}{\textit{Computation}} \\
$\Delta t_k$ & Variable time step size \\
$\mathbf{A}_k, \mathbf{X}_k$ & Discretized transition matrix and forcing term \\
$\bullet$ & Associative operator for parallel scan \\
\bottomrule
\end{tabular}
\end{table}

\subsection{Proof of Geometric Isometry and Gauge Invariance}
\label{app:geometry}
\textbf{Theorem 1.} \textit{Let $(\mathcal{M}, g)$ be a locally conformally flat Riemannian manifold with metric $g_{ij}(x) = e^{2\sigma(x)} \delta_{ij}$. If the encoder learns a gauge transformation $\phi(x) = 2\sigma(x)$, then the mapping $\Psi: T_x\mathcal{M} \to \mathbb{R}^d$ defined by the pointwise scaling $h = e^{\frac{1}{2}\phi(x)} v$ is an isometry.}
\begin{proof}
We consider a Clifford bundle $\mathcal{E} \to \mathcal{M}$ where each fiber is a Clifford algebra $Cl(T_x\mathcal{M})$. Let $v \in T_x\mathcal{M}$ be a tangent vector representing a physical quantity. The natural inner product on the tangent space induced by the physical Riemannian metric $g$ is

\begin{equation}
    \langle v, v \rangle_g = g_{ij} v^i v^j = e^{2\sigma(x)} \delta_{ij} v^i v^j = e^{2\sigma(x)} \|v\|_\delta^2,
\end{equation}

where $\|v\|_\delta$ is the standard Euclidean norm of the vector components in the local coordinate chart. The encoder $\mathcal{E}_\phi$ applies a learned local scaling operation modeled as a diagonal gauge transformation. In the pointwise limit, the latent variable is defined as $h = e^{\frac{1}{2}\phi(x)} v$. The squared Euclidean norm of this latent vector $h$ is

\begin{equation}
    \|h\|_\delta^2 = \langle e^{\frac{1}{2}\phi(x)} v, e^{\frac{1}{2}\phi(x)} v \rangle_\delta = \left(e^{\frac{1}{2}\phi(x)}\right)^2 \|v\|_\delta^2 = e^{\phi(x)} \|v\|_\delta^2.
\end{equation}

By minimizing the metric reconstruction loss $\mathcal{J}$, the network is forced to align the latent norm with the physical norm such that $\|h\|_\delta^2 \to \|v\|_g^2$. Equating the two energy expressions yields

\begin{equation}
    e^{\phi(x)} \|v\|_\delta^2 = e^{2\sigma(x)} \|v\|_\delta^2 \implies \phi(x) = 2\sigma(x).
\end{equation}

Thus, when the learnable parameter $\phi(x)$ converges to the conformal factor $2\sigma(x)$, the transformation preserves the metric structure. This implies that the flat latent space is isometric to the curved physical manifold enabling the use of Euclidean convolutions without geometric distortion.
\end{proof}

\subsection{Transient Growth in Non-Normal Systems}
\label{app:growth}
\textbf{Proposition 1.} \textit{For a linear dynamical system $\frac{dh}{dt} = \mathcal{L}h$, instantaneous energy growth $\frac{d}{dt}\|h\|^2 > 0$ is possible if and only if the numerical abscissa $\omega(\mathcal{L}) > 0$. In a stable system where the spectral abscissa $\alpha(\mathcal{L}) < 0$, such growth implies $\mathcal{L}$ must be non-normal.}
\begin{proof}
The time evolution of the system energy $E(t) = \|h(t)\|^2 = \langle h, h \rangle$ is given by

\begin{equation}
    \frac{dE}{dt} = \langle \dot{h}, h \rangle + \langle h, \dot{h} \rangle = \langle \mathcal{L}h, h \rangle + \langle h, \mathcal{L}h \rangle = \langle h, (\mathcal{L} + \mathcal{L}^\dagger)h \rangle.
\end{equation}

We define the numerical range of the operator $\mathcal{L}$ as the set of all Rayleigh quotients $W(\mathcal{L}) = \{ \langle \mathcal{L}x, x \rangle : \|x\|=1 \}$. The maximum possible instantaneous growth rate is bounded by the numerical abscissa $\omega(\mathcal{L})$

\begin{equation}
    \max_{h \neq 0} \frac{1}{\|h\|^2} \frac{dE}{dt} = \lambda_{\max}(\mathcal{L} + \mathcal{L}^\dagger) = 2\omega(\mathcal{L}).
\end{equation}

We distinguish two cases based on the normality of $\mathcal{L}$.
Case 1 Normal Operator. If $\mathbf{U} = \mathbf{V}$, then $\mathcal{L}$ commutes with its adjoint $[\mathcal{L}, \mathcal{L}^\dagger] = 0$. In this scenario, the numerical range is the convex hull of the spectrum. Thus $\omega(\mathcal{L}) = \alpha(\mathcal{L}) = \max \text{Re}(\lambda_i)$. Stability defined by $\alpha < 0$ necessitates strictly monotonic energy decay where $\omega < 0$.
Case 2 Non-Normal Operator. Since $\mathbf{U} \neq \mathbf{V}$ in our framework, $\mathcal{L}$ is non-normal. The numerical range can extend significantly into the right half plane even if all eigenvalues lie in the left half plane. It is algebraically possible to satisfy the condition $\alpha(\mathcal{L}) < 0 < \omega(\mathcal{L})$. This gap allows the system to extract energy from the non orthogonal eigenbasis resulting in transient growth before asymptotic decay. Therefore, UniPhy can model systems that are asymptotically stable yet exhibit transient energy amplification such as cyclogenesis.
\end{proof}

\subsection{Analytic Solution of the Stochastic Differential Equation}
\label{app:sde_derivation}
\textbf{Derivation.} We start with the continuous-time Stochastic Differential Equation governing the latent state

\begin{equation}
    dh_t = (\mathcal{L} h_t + \mathcal{R}) dt + \sigma dW_t.
\end{equation}

To solve this over the interval $[t_k, t_{k+1}]$ with $\Delta t = t_{k+1} - t_k$, we use the method of integrating factors. Multiplying by $e^{-\mathcal{L}t}$ yields

\begin{equation}
    d(e^{-\mathcal{L}t} h_t) = e^{-\mathcal{L}t} \mathcal{R} dt + e^{-\mathcal{L}t} \sigma dW_t.
\end{equation}

Integrating from $t_k$ to $t_{k+1}$ results in

\begin{equation}
    e^{-\mathcal{L}t_{k+1}} h_{k+1} - e^{-\mathcal{L}t_k} h_k = \int_{t_k}^{t_{k+1}} e^{-\mathcal{L}\tau} \mathcal{R} d\tau + \int_{t_k}^{t_{k+1}} e^{-\mathcal{L}\tau} \sigma dW_\tau.
\end{equation}

Multiplying both sides by $e^{\mathcal{L}t_{k+1}}$ provides the general solution

\begin{equation}
    h_{k+1} = \underbrace{e^{\mathcal{L}\Delta t}}_{\mathbf{A}_k} h_k + \underbrace{\int_{t_k}^{t_{k+1}} e^{\mathcal{L}(t_{k+1}-\tau)} \mathcal{R} d\tau}_{\mathbf{X}_{\text{drift}}} + \underbrace{\int_{t_k}^{t_{k+1}} e^{\mathcal{L}(t_{k+1}-\tau)} \sigma dW_\tau}_{\mathbf{X}_{\text{noise}}}.
\end{equation}

Assuming the nonlinear forcing $\mathcal{R}$ is constant over the small interval $\Delta t$, the drift term simplifies to

\begin{equation}
    \mathbf{X}_{\text{drift}} = \mathcal{R} \int_{0}^{\Delta t} e^{\mathcal{L}(\Delta t - u)} du = \mathcal{R} \mathcal{L}^{-1} (e^{\mathcal{L}\Delta t} - I).
\end{equation}

The noise term $\mathbf{X}_{\text{noise}}$ is a stochastic integral of a deterministic function with respect to the Wiener process. Its distribution is Gaussian with zero mean. To implement this efficiently, we must compute the covariance of this term. Using the Ito Isometry property $\mathbb{E}[(\int f(t) dW_t)^2] = \mathbb{E}[\int f(t)^2 dt]$, the variance $\mathbb{V}[\mathbf{X}_{\text{noise}}]$ is computed as

\begin{equation}
    \mathbb{V} = \mathbb{E} \left[ \left( \int_{0}^{\Delta t} e^{\mathcal{L}(\Delta t - \tau)} \sigma dW_\tau \right)^2 \right] = \sigma^2 \int_{0}^{\Delta t} \left| e^{\mathcal{L}(\Delta t - \tau)} \right|^2 d\tau.
\end{equation}

For the scalar diffusivity case assuming diagonality in the eigenbasis, this integral simplifies to

\begin{equation}
    \sigma^2 \int_{0}^{\Delta t} e^{2\text{Re}(\mathcal{L})(\Delta t - \tau)} d\tau = \sigma^2 \frac{e^{2\text{Re}(\mathcal{L})\Delta t} - I}{2\text{Re}(\mathcal{L})}.
\end{equation}

This analytic variance derivation allows us to sample the noise term $\epsilon_k$ in Algorithm \ref{alg:uniphy_forward} directly from a standard normal distribution scaled by this computed variance without performing expensive numerical stochastic integration.

\subsection{Associativity of the Affine Operator Semigroup}
\label{app:associativity}
\textbf{Theorem 2.} \textit{The binary operator $\bullet: \mathcal{S} \times \mathcal{S} \to \mathcal{S}$ defined as $(\mathbf{A}_j, \mathbf{X}_j) \bullet (\mathbf{A}_i, \mathbf{X}_i) = (\mathbf{A}_j \mathbf{A}_i, \mathbf{A}_j \mathbf{X}_i + \mathbf{X}_j)$ is associative.}
\begin{proof}
We can represent each affine transformation $(\mathbf{A}, \mathbf{X})$ acting on a state $h$ as a linear transformation in homogeneous coordinates. Let us define a matrix embedding $M_{(\mathbf{A}, \mathbf{X})}$ as

\begin{equation}
    M_{(\mathbf{A}, \mathbf{X})} = \begin{pmatrix} \mathbf{A} & \mathbf{X} \\ 0 & 1 \end{pmatrix}.
\end{equation}

The composition of two operators corresponds to matrix multiplication

\begin{equation}
    M_{(\mathbf{A}_j, \mathbf{X}_j)} M_{(\mathbf{A}_i, \mathbf{X}_i)} = \begin{pmatrix} \mathbf{A}_j & \mathbf{X}_j \\ 0 & 1 \end{pmatrix} \begin{pmatrix} \mathbf{A}_i & \mathbf{X}_i \\ 0 & 1 \end{pmatrix} = \begin{pmatrix} \mathbf{A}_j \mathbf{A}_i & \mathbf{A}_j \mathbf{X}_i + \mathbf{X}_j \\ 0 & 1 \end{pmatrix}.
\end{equation}

This matrix product exactly yields the definition of our operator $(\mathbf{A}_j \mathbf{A}_i, \mathbf{A}_j \mathbf{X}_i + \mathbf{X}_j)$. Since matrix multiplication is inherently associative, the operator $\bullet$ must also be associative. We verify this explicitly.

1. Left associative grouping
\begin{align}
    O_{32} = O_3 \bullet O_2 &= (\mathbf{A}_3 \mathbf{A}_2, \mathbf{A}_3 \mathbf{X}_2 + \mathbf{X}_3) \\
    O_{32} \bullet O_1 &= (\mathbf{A}_3 \mathbf{A}_2 \mathbf{A}_1, (\mathbf{A}_3 \mathbf{A}_2) \mathbf{X}_1 + (\mathbf{A}_3 \mathbf{X}_2 + \mathbf{X}_3)).
\end{align}

2. Right associative grouping
\begin{align}
    O_{21} = O_2 \bullet O_1 &= (\mathbf{A}_2 \mathbf{A}_1, \mathbf{A}_2 \mathbf{X}_1 + \mathbf{X}_2) \\
    O_3 \bullet O_{21} &= (\mathbf{A}_3 (\mathbf{A}_2 \mathbf{A}_1), \mathbf{A}_3 (\mathbf{A}_2 \mathbf{X}_1 + \mathbf{X}_2) + \mathbf{X}_3) \\
    &= (\mathbf{A}_3 \mathbf{A}_2 \mathbf{A}_1, \mathbf{A}_3 \mathbf{A}_2 \mathbf{X}_1 + \mathbf{A}_3 \mathbf{X}_2 + \mathbf{X}_3).
\end{align}

The two groupings yield identical results. Thus, the operator set forms a Monoid and validates the application of parallel scan algorithms.
\end{proof}

\subsection{Complexity Analysis of Parallel Scan}
\label{app:complexity}
\textbf{Theorem 3.} \textit{The parallel scan algorithm reduces the critical path for integrating a sequence of length $T$ from $O(T)$ to $O(\log T)$.}
\begin{proof}
Let the objective be to compute the prefix sums $S_k = O_k \bullet O_{k-1} \bullet \dots \bullet O_1$ for all $k \in [1, T]$. We distinguish between sequential scan and parallel scan.

Sequential Scan. The standard recurrence requires the output of step $k-1$ to compute step $k$. The dependency chain length is $T$. Total complexity is $\Theta(T)$.

Parallel Scan. We employ a balanced binary tree reduction.
First is the Up Sweep Phase where we compute pairwise compositions of adjacent elements. At depth $d$, we perform $T/2^{d+1}$ parallel operations. Since operations at the same depth are independent, the time complexity is proportional to the tree height given by $\log_2 T$.
Second is the Down Sweep Phase where we propagate the accumulated values from the root to the leaves to construct the prefixes. This also traverses the tree height adding another $\log_2 T$.

Conclusion. The total critical path length is $D(T) = 2\log_2 T = \Theta(\log T)$. While the total work remains $O(T)$, the exponential reduction in critical path allows for massive acceleration on parallel hardware compared to sequential processing.
\end{proof}

\section{Training Dynamics}
\label{app:training}

\begin{figure}[h]
    \centering
    \begin{minipage}{0.48\textwidth}
        \centering
        \includegraphics[width=\linewidth]{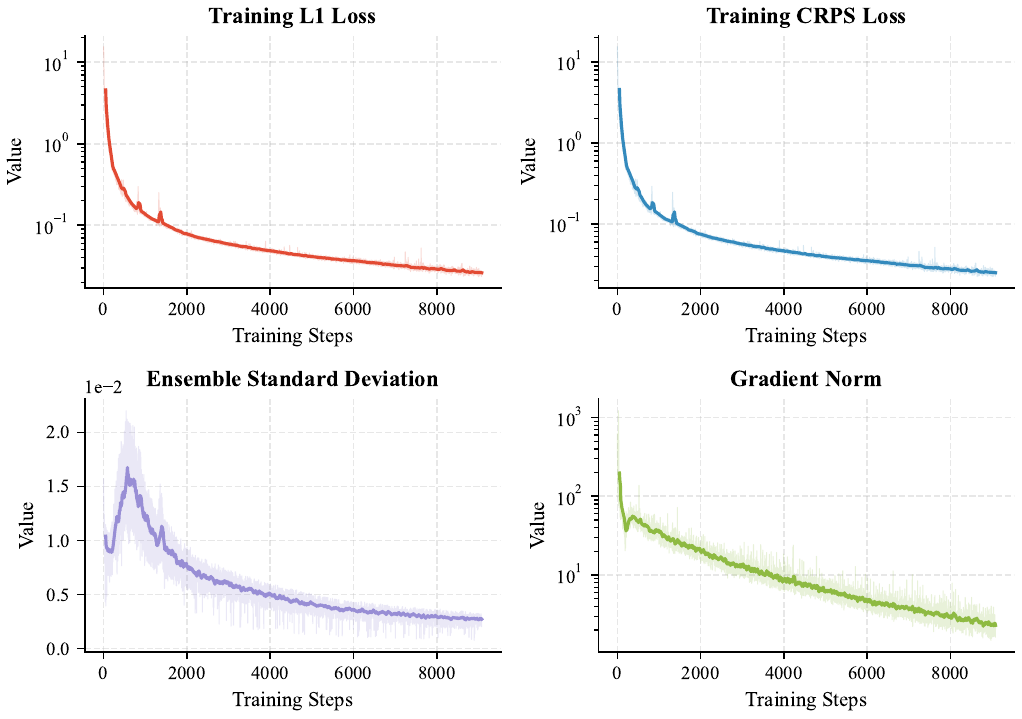}
        \caption{\textbf{Pre-training Dynamics.} Evolution of Loss, Ensemble Deviation, and Gradient Norm during the Neural SDE pre-training phase.}
        \label{fig:train_plots}
    \end{minipage}
    \hfill
    \begin{minipage}{0.48\textwidth}
        \centering
        \includegraphics[width=\linewidth]{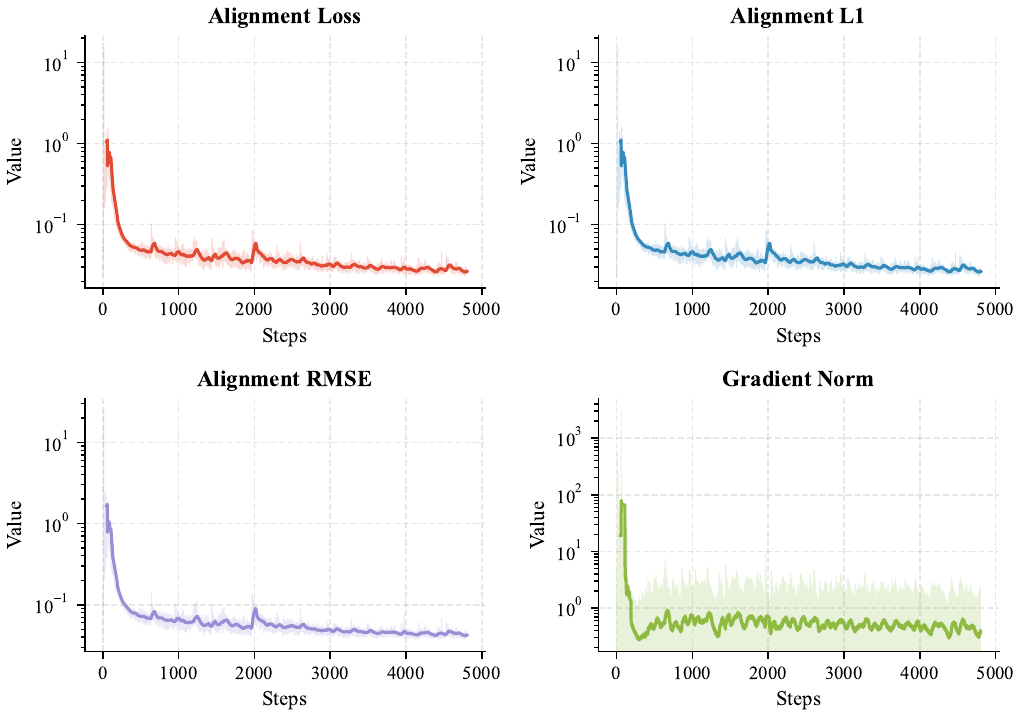}
        \caption{\textbf{Alignment Dynamics.} Convergence metrics during the fine-tuning phase, highlighting the reduction in long-term rollout error.}
        \label{fig:align_plots}
    \end{minipage}
\end{figure}

We monitor the stability of the training process by tracking the Loss (L1/CRPS), Gradient Norm, and Ensemble Standard Deviation. As shown in Figure \ref{fig:train_plots}, the pre-training stage (left) exhibits steady convergence over the course of training. The alignment stage (right) successfully reduces the multi-step rollout error without destabilizing the gradients, confirming the effectiveness of our two-stage optimization strategy.

\section{Hyperparameter Configurations}
\label{app:hyperparams}

Table \ref{tab:hyperparams} presents a side-by-side comparison of the specific hyperparameters used for the Pre-training (Stage 1) and Alignment (Stage 2) phases. This comparison highlights the distinct configurations, such as the change in gradient accumulation steps and embedding dimensions, tailored for each training objective.

\begin{table}[h]
    \centering
    \small
    \renewcommand{\arraystretch}{1.2}
    \caption{\textbf{Hyperparameter Comparison: Pre-training vs. Alignment.}}
    \label{tab:hyperparams}
    \setlength{\tabcolsep}{8pt}
    \begin{tabular}{l|l|c|c}
        \toprule
        \textbf{Category} & \textbf{Parameter} & \textbf{Pre-training} & \textbf{Alignment} \\
        \midrule
        \multirow{7}{*}{\textbf{Model}} 
        & Embedding Dimension & 128 & 512 \\
        & Layers (Depth) & 8 & 8 \\
        & Expansion Ratio & 4 & 4 \\
        & Patch Size & $(7, 15)$ & $(7, 15)$ \\
        & Image Size & $721 \times 1440$ & $721 \times 1440$ \\
        & Ensemble Size & 4 & 4 \\
        & Max Growth Rate & 0.3 & 0.3 \\
        \midrule
        \multirow{4}{*}{\textbf{Data}} 
        & Sampling Mode & \texttt{mix} & \texttt{sequential} \\
        & Window Size & 10 & 10 \\
        & Look Ahead & 2 & - \\
        & Reference $\Delta t$ & 6.0 h & 6.0 h \\
        \midrule
        \multirow{8}{*}{\textbf{Optimization}} 
        & Learning Rate & $1.0 \times 10^{-4}$ & $5.0 \times 10^{-5}$ \\
        & Weight Decay & 0.05 & 0.01 \\
        & Epochs & 256 & 20 \\
        & Batch Size & 1 & 1 \\
        & Gradient Clipping & 128.0 & 10.0 \\
        & Grad Accumulation & 1 & 4 \\
        & SDE Mode & \texttt{"sde"} & \texttt{"sde"} \\
        \midrule
        \multirow{3}{*}{\textbf{Alignment}} 
        & Condition Steps & - & 4 \\
        & Max Target Steps & - & 2 \\
        & Sub-steps & - & $[1, 2, 3, 6]$ \\
        \bottomrule
    \end{tabular}
\end{table}

\end{document}